# The effect of dissipative corona on the structure and stability of cold optically thick accretion disks at high accretion rates

Ranjeev Misra[1]

and

Ronald E. Taam[2]

## ABSTRACT

The vertical structure of optically thick accretion disks is investigated in the two-zone approximation. The disk is divided into an underlying disk and a corona, where the latter is defined as the upper surface layers for which the effective optical depth is unity. It is found that a significant part of the accretion flow (or dissipation rate) can take place in the corona if the scale height of the magnetic field is larger than that of the disk. The presence of such a dissipative corona leads to a modification in the topology of local disk solutions. For example, these solutions are found from local stability analysis to be both secularly and thermally stable, for accretion rates which are a factor $\approx$ four higher than those inferred from the stability of standard disk solutions. Thus, the applicability of optically thick disks with dissipative coronas are not as restrictive as disks without such coronas and can provide an attractive explanation for the origin of the soft spectral component observed in black hole X-ray binary systems.

*Subject headings:* accretion, accretion disks-black hole physics -hydrodynamics

## 1. Introduction

X-ray binary systems in which one star is believed to be a black hole are primarily, but not solely, observed in one of two long term spectral states. In the so called soft state, the spectrum of the system consists of a soft black body-like component (kT $\approx$ 1 keV), which often dominates the luminosity, and a hard power-law tail with a photon index $\Gamma \approx 2.5$. In contrast, the hard spectral state can be described approximately as a harder power-law ($\Gamma \approx 1.7$) while the soft component is either weak or undetectable. Although the existence of these spectral states has been known for a long time, there is no general consensus for their origin. In particular, several models have been proposed for the possible origin of the hard X-ray power-law. For example, the emission could arise from thermal Comptonization of soft photons in a hot accretion disk (Shapiro, Lightman & Eardley 1976; Esin, McClintock & Narayan 1997) or from bulk Comptonization in a converging flow (Laurent & Titarchuk 2001). Other scenarios invoke a hot low density Comptonizing coronal region sandwiching a cold accretion disk (Liang & Price 1977; Haardt & Maraschi 1993). Within the framework of these models, active time varying regions (or flares) on top of the disk may be responsible for the production of the hard X-ray spectrum (Poutanen & Fabian 1999). In yet another model, the hard X-rays may arise from a disk where the temperature is a rapidly varying function of radius (Misra & Melia 1996). Although these models are characterized by different disk structure and geometry, they all attribute the low energy component of the soft state to emission from a cold accretion disk. Thus, there seems to be a general consensus that, at least, during the soft state, a cold accretion disk exists extending to nearly the innermost stable orbit in these systems.

[2]Inter-University Center for Astronomy and Astrophysics, Ghaneskhind Road, Pune, India; rmisra@iucaa.ernet.in

[2]Department of Physics and Astronomy, Northwestern University, 2131 Sheridan Road, Evanston,IL 60208; taam@apollo.astro.nwu.edu

In the standard theory of accretion disks pioneered by Shakura & Sunyaev (1973), the turbulent viscosity is parameterized as

$$\tau_{r\phi} = \alpha P, \tag{1}$$

where $\tau_{r\phi}$ is the viscous torque, $P$ is the pressure and $\alpha$ is the viscosity parameter. A cold optically thick disk with such a viscosity prescription is known to be secularly and thermally unstable if in the disk the ratio of the gas pressure to the total pressure is less than 0.4 (Lightman & Eardley 1974; Shakura & Sunyaev 1976). This occurs, for example at a radius, R, of $20GM/c^2$ when the accretion rate is

$$\dot{M} > (10^{17} \text{g s}^{-1})\alpha^{-1/8}(\frac{M}{10M_\odot})^{7/8} \tag{2}$$

which corresponds to luminosities greater than $5\times10^{36}$ ergs s$^{-1}$ for a ten solar mass black hole. Although several X-ray binaries exceed this luminosity in the soft state, they generally do not exhibit large amplitude temporal variations which are expected if the secular and thermal instabilities are indeed present (Taam & Lin 1984). An exception to this statement is the microquasar GRS 1915+105 (Mirabel & Rodriguez 1999) where, indeed, large amplitude variations are often observed. It should be emphasized that these predicted instabilities depend on the form of the viscosity prescription and occur when the viscosity and the temporal response of the viscosity is highly temperature sensitive (e.g., Honma, Matsumoto, & Kato 1991).

A promising mechanism for driving the turbulence responsible for angular momentum and energy transport is the action of the magneto-rotational instability (MRI) that is expected to take place in such disks (Balbus & Hawley 1991). In fact, recent 3D magneto hydrodynamic (MHD) simulations have shown that the MRI can give rise to a turbulent viscosity which leads to accretion flow in a Keplerian disk (Hawley, Balbus & Stone 2001). The time averaged viscous torque in these simulations was found to be proportional to the magnetic pressure which itself is locally proportional to the disk pressure. Thus, the form of eqn (1), as a physical viscosity prescription was justified. However, these simulations do not include the effect of radiation and, hence, it is not clear whether this prescription is valid when the disk is radiation pressure dominated. On the other hand, there are no definite reasons to believe that it will be different. In this framework, eqn (1) may well be valid only in a time averaged sense. In this case, the characteristic timescale ($t_{ave}$) over which the averaging needs to be performed can only be estimated after MHD simulations with radiation and cooling are undertaken. Any inferred temporal response of the disk using eqn (1) on timescales shorter than $t_{ave}$ can be misleading. Hence, the thermal instability inferred from the use of eqn (1) may not occur if the thermal timescale is much shorter than $t_{ave}$. However, the secular instability occurs on the viscous timescale which should be longer or at least comparable to $t_{ave}$. Thus, contrary to observations, radiation pressure dominated disks are expected to be non steady, suffering from, at least, the secular instability.

We point out that these predictions of instability are based on simplifying assumptions on the vertical distribution of the energy dissipation rate in the disk. In particular, it is possible that the vertical disk structure is sufficiently complex that these instabilities may be suppressed. For example, the existence of a corona in the upper layers of the disk could play a dynamically and energetically important role in the accretion process. Such a dissipative corona (i.e., a corona where a substantial fraction of the energy dissipated and/or accretion flow takes place) may be able to stabilize the underlying disk (see Ionson & Kuperus 1984; Svensson & Zdziarski 1994; Chen 1995; Abramowicz, Chen, & Taam 1995; Zycki, Collin-Souffrin, & Czerny 1995). To compute the vertical structure of the disk, a knowledge of the energy dissipation rate as a function of height is required. Due to the uncertainties involved the dissipation rate has been assumed, in the literature, to be proportional to the gas density (e.g., Ross & Fabian 1996, Shimura & Takahara 1995). With this assumption, the dissipation in the low density upper layers of the disk is negligible compared to its contributions in the midplane. However, the dissipation rate is proportional to the magnetic pressure for the MRI induced viscosity, and numerical simulations show that the magnetic pressure scale height is larger than the disk scale height (Hawley, Balbus & Stone 2001). In such a description, a low density but highly magnetized dissipative corona can form which may act to stabilize the disk.

In this work, we investigate the conditions under which a dynamically important dissipative corona exists and study the effect of such a corona on the stability of the disk. A simplified approach is adopted where the disk is divided into two distinct regions: the underlying disk and the corona. Here, the corona is explicitly defined as the upper layer of the disk where the effective optical depth is unity. Thus the definition of a corona here differs from previous works where an optically thin distinct region above an optically thick disk is referred to as the corona. In this approach, the disk structure equations are reduced to simple algebraic relations in terms of average quantities of the disk and corona. Such an analysis, while giving qualitative results, is more suitable for comparison with the standard vertically integrated disk solutions and stability analysis.

In the next section, the disk and corona structure equations are formulated and the assumptions underlying our analysis are presented. The solutions of these equations are described in §3, and the stability of the disk is analyzed in §4. Finally, we summarize our work and discuss its implications in the last section.

## 2. Disk Structure Equations

The disk and overlying corona structures are cast into algebraic form in a development similar to standard thin disk theory. In contrast to the standard theory, the effect of the corona on the disk and vice versa are explicitly taken into account. The following approximate equations describe the vertically averaged quantities of the disk and corona. Throughout this paper, quantities with D (C) subscript denote the disk (corona).

Hydrostatic equilibrium for the disk gives,

$$\frac{P_D - P_C}{H_D} = \frac{GMm_p}{R^3} n_D H_D \tag{3}$$

while for the corona it is given by,

$$\frac{P_C}{H_C} = \frac{GMm_p}{R^3} n_C (H_C + H_D) \tag{4}$$

where $P$ is the total pressure, $R$ is the radius, $H$ is the height, $n$ is the number density and $M$ is the mass of the black hole. The total pressure consists of contributions from gas, radiation and magnetic pressures. In the above, we have assumed that the disk and corona are geometrically thin (i.e., $H << R$). Thus for the disk,

$$P_D = P_{DR} + P_{DG} + P_{DB} = 2n_D k T_D + 4\sigma T_D^4/3c + P_{DB} \tag{5}$$

and similarly for the corona,

$$P_C = P_{CR} + P_{CG} + P_{CB} = 2n_C k T_C + 4\sigma T_C^4/3c + P_{CB} \tag{6}$$

where $T$ is the temperature and $P_{DB}$ ($P_{CB}$) is the magnetic pressure in the disk (corona). For simplicity, the gas is assumed to be composed of pure hydrogen. It can be seen that the effect of the corona on the underlying disk enters through eqn (3) in the matching pressure condition at the interface of the disk and corona.

The viscous torque is assumed to be proportional to the magnetic pressure. Hence the energy flux dissipated in the disk is,

$$F_{DG} = \frac{3}{2}\alpha_B P_{DB} H_D \Omega_K \tag{7}$$

while for the corona it is,

$$F_{CG} = \frac{3}{2}\alpha_B P_{CB} H_C \Omega_K \tag{8}$$

where $\Omega_K$ is the Keplerian angular velocity and the constant of proportionality $\alpha_B$ is assumed to be same for both the disk and the corona. For the disk, the dissipated flux is matched by the radiative flux,

$$F_{DR} = \frac{4\sigma}{3\tau_D}(T_D^4 - T_C^4) = F_{DG} \tag{9}$$

For the corona the entire flux energy generated in the disk is radiated away. Hence,

$$F_{CR} = \frac{4\sigma T_C^4}{3\tau_C} = F_{CG} + F_{DG} \tag{10}$$

By definition the effective optical depth in the corona is unity,

$$\tau_{C*}^2 = \tau_C \times 1.8 \times 10^{-25} n_C^2 T_C^{-7/2} H_C = 1 \tag{11}$$

Note that since $\tau_{C*} = 1$, the corona is expected to be in approximate thermal equilibrium (see for e.g. Ross & Fabian 1996) which justifies eqn (10) and the expression for radiation pressure in the corona, taken to be equal to $4\sigma T_C^4/3c$. Essentially, although the Compton y-parameter in the corona is large, the free-free absorption is still dominant, making the radiation nearly thermal.

It is assumed here that the magnetic pressure in the disk is a fraction of the gas and radiation pressure. Thus

$$P_{DB} = \beta_D(P_{DR} + P_{DG}) \tag{12}$$

where $\beta_D$ measures the fraction of the magnetic pressure to its equipartition value in the disk. To take account of the possibility that the scale height of the magnetic field could differ from that of the density, the magnetic pressure in the corona is assumed to be given by,

$$P_{CB} = P_{DB}[\frac{(P_{CR} + P_{CG})}{(P_{DR} + P_{DG})}]^{\gamma} \qquad (13)$$

Note that if the exponent $\gamma = 1$, the ratio of the magnetic pressure to its equipartition value is the same in the disk as well as in the corona. At the other extreme if $\gamma = 0$, the magnetic field is the same for the disk and the corona which corresponds to the limit where the magnetic scale height is much larger than that of the disk. This implies that the magnetic field in the corona exceeds its equipartition value (Hawley, Balbus & Stone 2001).

The disk and corona structure can now be solved based on eqns (3-13) for various values of the total optical depth $\tau_T = \tau_D + \tau_C$ for a given mass of the black hole $M$, the distance $R$ from the black hole, the proportionality constant $\alpha_B$, disk equipartition fraction $\beta_D$ and the exponent $\gamma$. For each solution the accretion rates in the disk and corona can be inferred from

$$F_{DG} = \frac{3}{8\pi} \frac{GM\dot{M}_D}{R^3} \qquad (14)$$

$$F_{CG} = \frac{3}{8\pi} \frac{GM\dot{M}_C}{R^3} \qquad (15)$$

where the boundary effects have been neglected. If a zero-torque boundary condition in the last stable orbit is enforced then the accretion rates would be a factor $1/J$ larger than the ones inferred using Eqns (14) and (15), where $J = 1 - (6GM/c^2R)^{1/2}$ for a black hole described in terms of a Schwarzschild metric.

## 3. Results

The topology of the local disk solutions in $(\dot{M}_T, \tau_T)$ space is shown in Fig. 1 for a given black hole mass, radius, $\alpha_B$, $\beta_D$ and for various values of $\gamma$. Here $\dot{M}_T$ is the total accretion rate (i.e., the sum of the mass flow rate through the corona and the disk). In Figures 2 (a-f), the ratios of structural parameters of the disk and corona are shown as a function of $\dot{M}_T$. For reference the well known topology of the solutions for the standard accretion disk theory for the same set of parameters is also shown in Fig. 1 (dashed line). For small accretion rates ($\dot{M} < 10^{17}$ g s$^{-1}$), the disk solutions are gas pressure dominated and $\dot{M}_T$ increases with $\tau_T$, while for higher accretion rates, the disk is radiation pressure dominated and $\dot{M}_T$ decreases with $\tau_T$. Here the solutions have been terminated when the disk becomes optically thin and the assumption that the radiation is in thermal equilibrium breaks down. Note that the solutions for disks with dissipative coronas require one more parameter ($\gamma$) than the standard disk model, which characterizes the variation of the scale height of the magnetic field as compared to the disk scale height. The set of solutions for $\gamma = 1$, has not been shown since they are similar to the standard disk ones.

For small accretion rates $\dot{M}_T < 10^{17}$ g s$^{-1}$, the coronal accretion rate is much smaller than within the disk (except for the case when $\gamma = 0$) and hence the effect of the corona on the underlying disk is minimal. The solutions are similar to models of the standard disk. At high $\dot{M}_T$, the solutions are different than the standard ones. In particular, the accretion rate at which the slope of $\dot{M}_T$ vs $\tau_T$ changes sign, is higher. For $\gamma = 0$, this slope at high accretion rates is steeper than that for standard disk solutions. Fig 2 (b) shows that for low values of $\gamma < 0.25$, $\dot{M}_C \approx \dot{M}_D$, and hence it for these values of $\gamma$ that a dissipative corona can exist. Figures (3) and (4) show how the solutions change when $\beta_D$ and the radius is varied. In particular, the critical mass accretion rate above which the slope of $\dot{M}_T$ vs $\tau$ changes sign increases with decreasing values of $\beta_D$ and increasing radius $R$. Varying $\alpha_B$, has a similar effect as changing $\beta_D$.

Simple analytical expressions for the disk structure can be obtained for the standard disk model when either the disk is radiation or gas pressure dominated. For disks with dissipative corona the situation is more complex. Several different limits and their combination can be considered. For example, the disk could be radiation or gas pressure dominated while the corresponding corona could be gas or radiation pressure dominated. Figs. 2 (e) and 2 (f) show that, in general, all four possible combinations are realizable. Moreover, to obtain analytical expressions, one has to assume that either the accretion rate through the corona or the disk dominates over the other. While it is possible to present such analytical expressions, they are not generally valid for the most interesting solutions (typically at high accretion rates). Thus, only the numerical solutions to the disk structure equations have been presented.

In this work, the emergent spectra from such disks have not been computed, since a detailed radiative transfer calculation in the vertical direction would be required. The corona is, by definition, marginally optically thick (i.e. $\tau_{C*} = 1$), and hence the radiation is expected to be nearly in thermal equilibrium. The emergent spectrum should be a black body (with flux proportional to the total dissipation) diluted by the effect of Comptonization in the uppermost layers of the corona where the Compton y-parameter is of order unity (Ross & Fabian 1996). Thus, the total spectrum from the disk should be close to the multi-color disk spectrum which is often used for fitting the soft spectral component of black hole X-ray binary systems.

## 4. Stability of disks with dissipative coronas

In the standard accretion disk theory, radiation pressure dominated disks are known to be both secularly and thermally unstable. The secular instability criterion depends on the topology of solutions in $(\dot{M}, \tau)$ space, where if $\dot{M}$ is inversely proportional to $\tau$, the disk would be locally unstable. Thus, whether disks with dissipative coronas are secularly stable or not, can be directly inferred from Fig. 1. For $\gamma < 0.5$, which correspond to nearly the same equipartition fraction for the disk and corona, the accretion rate at which the disk is locally unstable (Fig 1, curves 1) is nearly same as that for the standard disk theory. On the other hand, when $\gamma = 0$ (Fig. 1, curve 3), a secularly stable disk can exist for $\dot{M} < 4 \times 10^{17}$ g s$^{-1}$ which is $\approx 4$ higher than what is expected for the standard disk solution (eqn 1). Moreover, for $\gamma = 0$, in the unstable branch, the dependence of $\dot{M}_T$ on $\tau_T$ is steeper and for a given accretion rate the steady state optical depth is larger than that predicted by the standard theory. This means that the secular instability would occur on a longer time-scale and the global evolution of such disks could be significantly different than the scenario predicted by time dependent numerical simulations of standard disks (Taam & Lin 1984). Thus disks with dissipative coronas, with low values of $\gamma$ (i.e., when the scale height of the magnetic field is sufficiently large compared to the disk scale height) may be secularly stable for higher accretion rates than standard disks.

While the secular instability occurs on the viscous timescale, the thermal instability occurs on the much shorter thermal timescale. As discussed in §1, a thermal stability analysis using the viscosity law (eqn. 1) is valid only when the averaging timescale of the viscous law is shorter than the thermal timescale. Thus, the thermal stability analysis presented here may not be directly applicable to disks with dissipative coronas. Given this caveat, for the total disk to be thermally stable, both the underlying disk and the corona must be stable individually. The thermal stability criterion for the disk is,

$$\xi_{DR} > \xi_{DG} \tag{16}$$

where,

$$\xi_{DG} = \frac{d \log F_{DG}}{d \log T_D}\Big|_{\tau_T} \tag{17}$$

and

$$\xi_{DR} = \frac{d \log F_{DR}}{d \log T_D}\Big|_{\tau_T} \tag{18}$$

Here, the derivatives are taken while maintaining the total optical depth, $\tau_T$, to be constant. Since the thermal timescale of the corona is much smaller than in the disk, it is assumed here that the corona is in thermal equilibrium during the perturbation. The stability criterion for the corona is,

$$\xi_{CR} > \xi_{CG} \tag{19}$$

where,

$$\xi_{CG} = \frac{d \log F_{CG}}{d \log T_C}\Big|_{P_D} \tag{20}$$

and

$$\xi_{CR} = \frac{d \log F_{CR}}{d \log T_C}\Big|_{P_D} \tag{21}$$

Here, the derivatives are taken with the underlying disk pressure, $P_D$, held constant since the thermal timescale of the disk is much larger than that of the corona. These logarithmic derivatives are numerically computed and shown in Figs. 5 (a) and (b). For $\gamma = 0$, Fig 5 (a) shows that $\xi_{CR}$ (solid line) is always greater than $\xi_{CG}$ (dashed line). Hence the corona is thermally stable even when it is radiation pressure dominated (Fig. 2 f). The underlying disk becomes unstable when radiation pressure dominates and hence $\xi_{DR}$ is less than $\xi_{DG}$ (Fig. 5 b) for $\dot{M}_T > 4 \times 10^{17}$ g s$^{-1}$. Thus, disks with dissipative coronas with low values of $\gamma$, are both thermally and secularly stable for accretion rates less than about $4 \times 10^{17}$ g s$^{-1}$ (for parameter values used in Fig 1) corresponding to a luminosity of $\approx 2 \times 10^{37}$ ergs/sec or 0.01 times the Eddington Luminosity for a 10 solar mass black hole.

## 5. Summary and Discussion

In this study we have investigated the vertical disk structure in the two zone approximation where the disk is divided into a disk and an overlying corona. In contrast to previous studies, the corona is defined as the upper regions of the disk with an effective optical depth of unity. It is shown that if the scale height of the magnetic field in the disk is larger than the density scale height, a dissipative corona can form where significant accretion flow (and hence significant dissipation) takes place. The topology of such disk solutions differs from that of the standard disk solutions. For the case when the magnetic scale height is much greater than the disk scale height (i.e. when the magnetic field is constant in the vertical direction corresponding to $\gamma = 0$), the disk is both thermally and secularly stable for accretion rates less than $\dot{M} = 4 \times 10^{17}$ g s$^{-1}$ ( for a ten solar mass black hole and $R = 20GM/c^2$) as compared to $\dot{M} = 10^{17}$ g s$^{-1}$ for the standard disk. Although this accretion rate is nearly independent of the viscosity parameterization, $\dot{M} \propto \beta^{0.2}$ (Fig. 3), it is possible that it may be higher if either a) the magnetic field in the corona is larger than that in the disk (i.e.

$\gamma < 0$) or b) if a significant amount of the energy dissipated in the disk is transported non-radiatively to the corona. We defer such detailed and perhaps more speculative studies for a future work. Thus, disks with dissipative coronas may provide a more attractive explanation for the source of the high luminosity soft spectral component observed in black hole X-ray binary systems.

The two zone approximation employed in this study, simplifies the disk equations to algebraic form and provides qualitative insight into the existence and effect of dissipative coronas. Clearly, there is a need to confirm these results and to obtain quantitative solutions by constructing models based on calculations of the vertical radiative transfer differential equations. The simple analysis undertaken here points not only to the possible existence of dissipative coronas but also provides a description of qualitative different boundary conditions required for the solution of the radiative transfer equations. The complete radiative transfer analysis will also yield theoretical emergent spectra from such disks which can then be compared with observations of black hole X-ray binary systems. Such meaningful comparisons with observations would provide a test for the hypothesis that the soft X-ray emission originates from disks with dissipative coronas

We thank the referee for the detailed comments which has significantly improved the manuscript. R.M. gratefully acknowledges support from the Lindheimer Fellowship at Northwestern University.

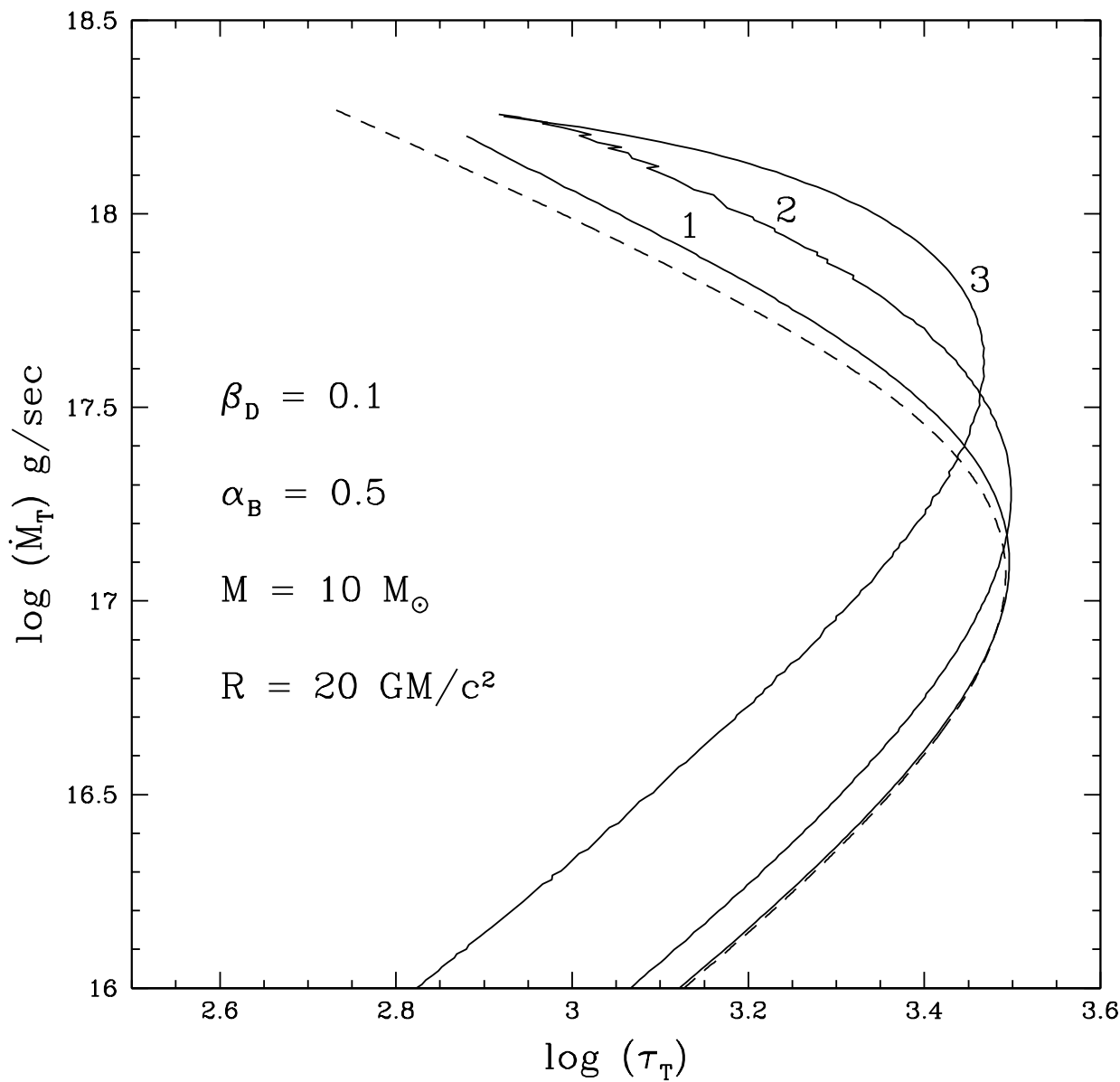

Fig. 1.— Total accretion rate ($\dot{M}_T = \dot{M}_C + \dot{M}_D$) versus total optical depth ($\tau_T$). Curve 1: $\gamma = 0.5$. Curve 2: $\gamma = 0.25$. Curve 3: $\gamma = 0$. Dashed line is for the standard accretion disk solutions. The solutions have been terminated when the effective optical depth of the disk becomes inconsistently less than unity.

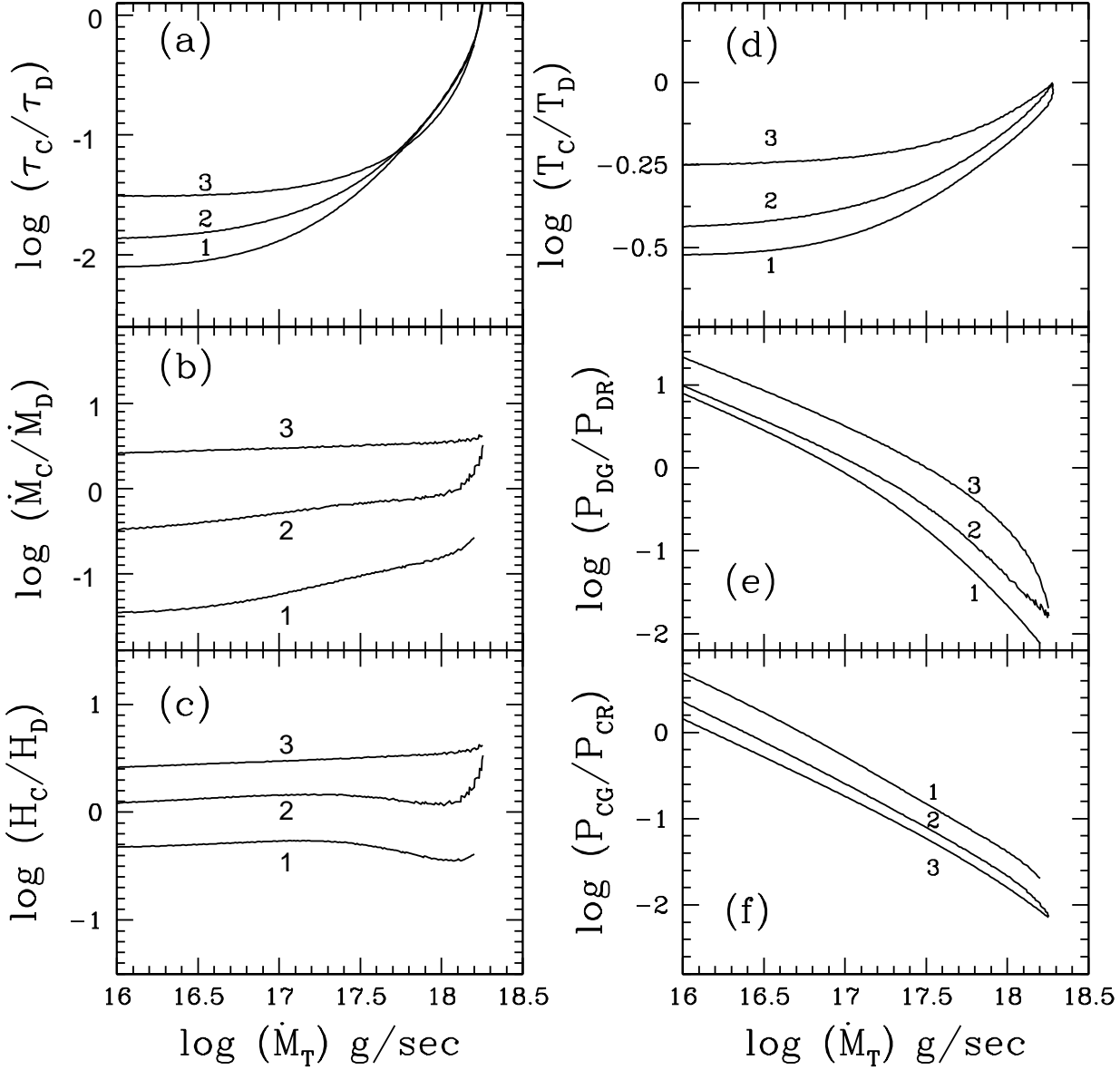

Fig. 2.— Variation of various ratios with accretion rate. Parameters and curve numbers ( corresponding to different values of $\gamma$) are same as in Fig 1. Note that in general $\tau_D >> \tau_C$ and for $\gamma < 0.25$ the coronal accretion rate is always comparable to the disk accretion rate.

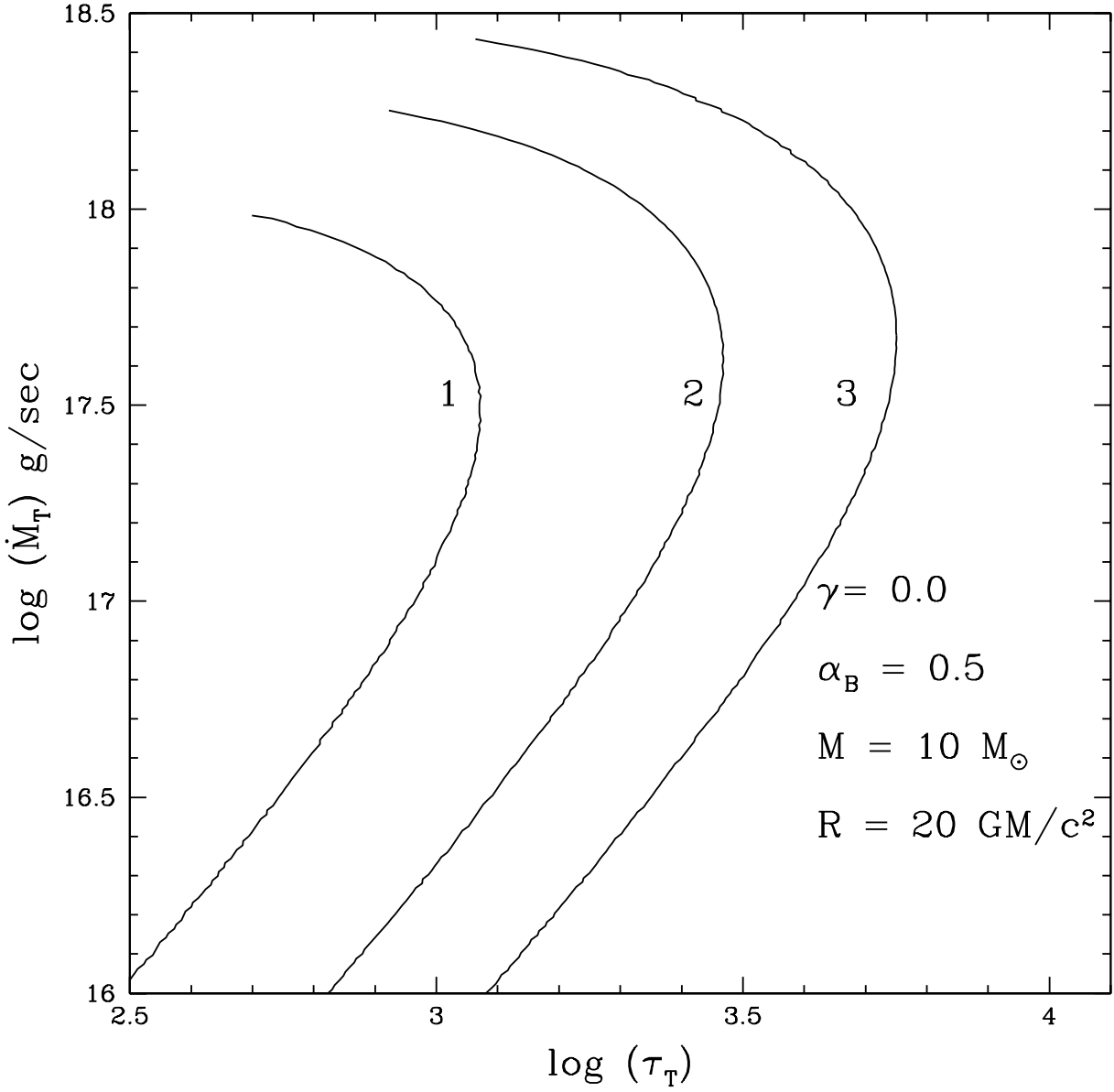

Fig. 3.— Total accretion rate ($\dot{M}_T = \dot{M}_C + \dot{M}_D$) versus total optical depth ($\tau_T$). Curve 1: $\beta_D = 0.25$. Curve 2: $\beta_D = 0.1$. Curve 3: $\beta_D = 0.05$.

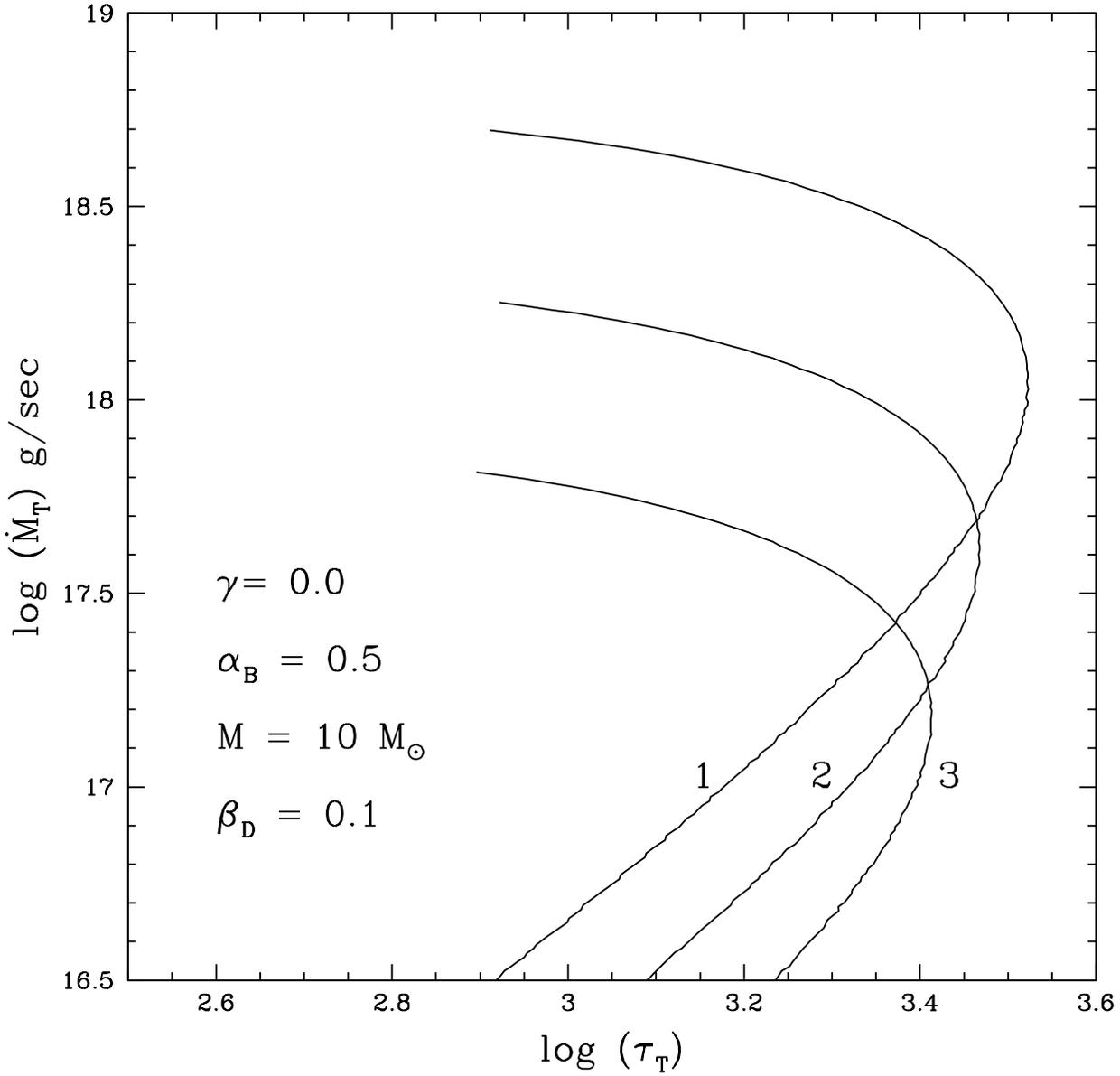

Fig. 4.— Total accretion rate ($\dot{M}_T = \dot{M}_C + \dot{M}_D$) versus total optical depth ($\tau_T$). Curve 1: $R = 40GM/c^2$. Curve 2: $R = 20GM/c^2$. Curve 3: $R = 10GM/c^2$.

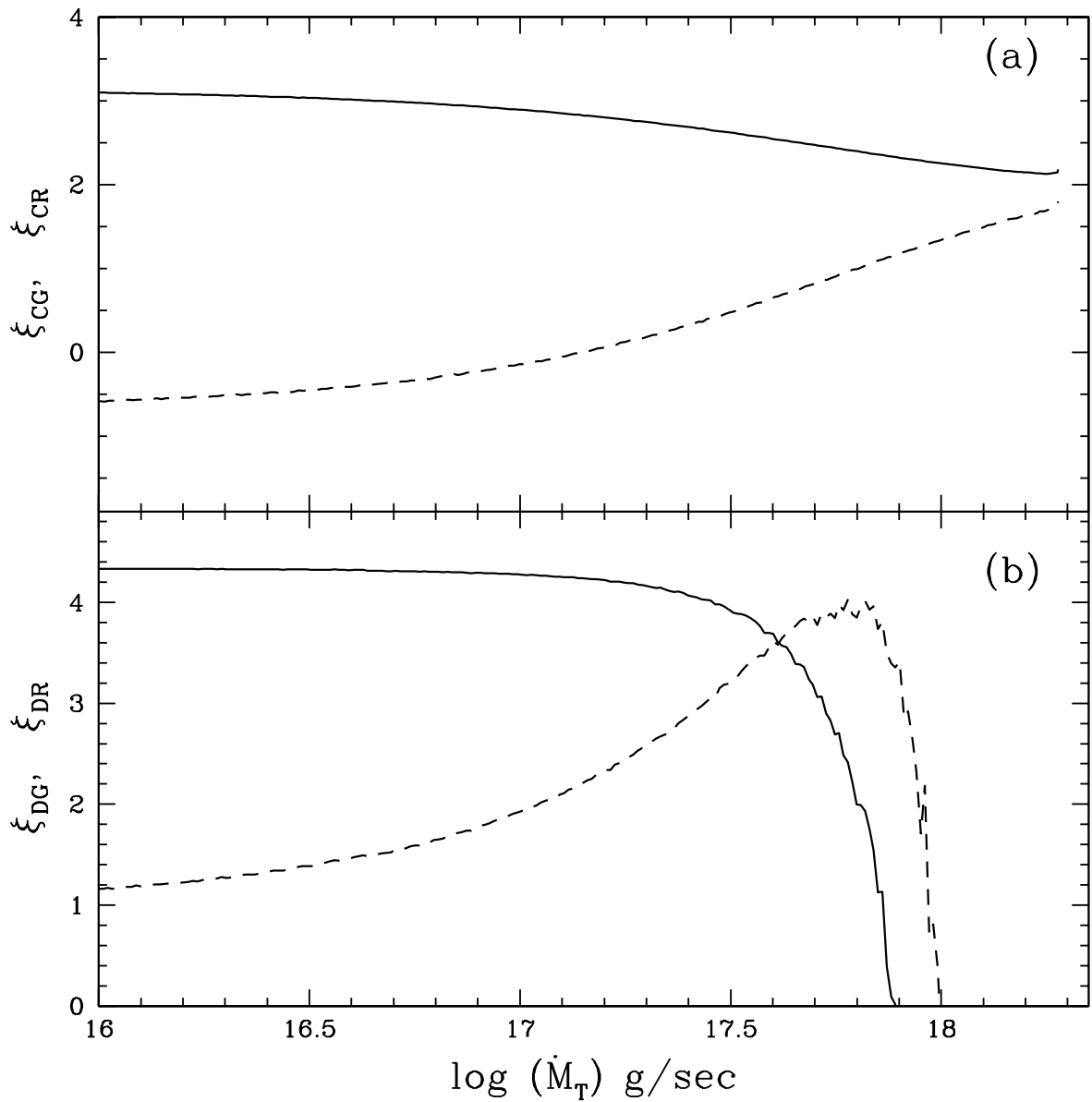

Fig. 5.— (a) $\xi_{CG}$ (dashed line) and $\xi_{CR}$ (solid line) versus accretion rate for $\gamma = 0$ (curve 3 of Figure 1). The corona is thermally stable. (b) $\xi_{DG}$ (dashed line) and $\xi_{DR}$ (solid line) versus accretion rate for $\gamma = 0$ (curve 3 of Figure 1). The disk is thermally stable for accretion rate less $4 \times 10^{17}$ g/sec.